# Design of a New Stream Cipher: PALS

*Mohammadreza Ashouri , University of Potsdam, Germany*

*Ashouri@uni-potsdam.de*

*Abstract*— In this paper, a new stream cipher is designed as a clock-controlled one, but with a new mechanism of altering steps based on system theory in such a way that the structures used in it are resistant to conventional attacks. Our proposed algorithm (PALS) uses the main key with the length of 256 bits and a 32-bit message key. The most important criteria considered in designing the PALS are resistance to known attacks, maximum period, high linear complexity, and good statistical properties. As a result, the output keystream is very similar to the perfectly random sequences and resistant to conventional attacks such as correlation attacks, algebraic attack, divide & conquer attack, time-memory tradeoff attack and AIDA/cube attacks. The base structure of the PALS is a clock-controlled combination generator with memory and we obtained all the features according to design criteria with this structure. PALS can be used in many applications, especially in financial cryptography due to its proper security features.

## I. INTRODUCTION

New stream ciphers have been designed for over four decades and never stopped. The reason is the need for new algorithms in different platforms, for organizations and governments and for updating and proving their resistance to various new attacks. Therefore, cryptographers always try to design new algorithms from different perspectives; some for higher security, some for more speed and some for utilizing fewer resources. Thus, the necessity of designing new algorithms is quite obvious.

Stream cipher cryptosystems are one of the most important data encryption systems which have found significant application in strategic sectors. In a stream cipher system, plaintext digits are combined with a pseudorandom keystream to produce ciphertext [1]. If the keystream is a random sequence with independent components and uniform distribution, the ciphertext will not give any information to the attacker about the plaintext. In other words, the mutual information between the plain text and cipher text is zero and the system has complete security. Vernam or one-time pad system is the only system that works on this basis. Due to practical problems, this system is used in certain cases. Today, of course, due to the advancement of technology and the availability of ultra-large memory in small volumes, the practical use of this system has been facilitated.

In practical systems, a pseudo-random generator is usually used to generate the keystream. The generator should be designed very similar to a BSS[1] generator. In other words, the keystream should be the same as the perfectly random sequence. These generators produce sequences with desirable statistical properties, such that it is impossible to achieve the main key from keystream. In general, the design approaches for keystream generators can be categorized into four categories [2, 3]:
- Information theory
- Complexity theory
- System theory
- Provably secure systems

In the first approach, the design basis is mutual information between the plaintext and the cipher text. A cryptosystem is secure from the perspective of information theory, whenever the attackers cannot obtain any information about the probability distribution of possible messages in the cipher-text-only attack, despite their unlimited computational power.

In the second approach, the design basis is the use of one-way functions. The main goal of the complexity theory is that the enemy with limited power cannot distinguish the keystream from a truly random sequence. Such a generator is called computationally secure. In order to implement this method, various criteria have been presented so far.

In the system theory approach, the focus is on popular and well-known attacks. In this way, the design of a system is done in such a way that it is resistant to those attacks. A cryptosystem is called practically secure from the perspective of system theory, whenever it is not breakable by any known attack in a reasonable time.

In the fourth approach, all attacks against the system are considered, with the difference that in this approach, a cryptosystem is provably secure, whenever a lower bound for the average computations required in each attack can be proved. It is also known as the unconditional security.

Our proposed algorithm is designed based on system theory and the structures used in it are resistant to known attacks.

## II. DESIGN CRITERIA

The most important criteria considered in designing the proposed algorithm are:
- Maximum period
- High linear complexity
- Resistance to known attacks (correlation attacks, algebraic and time-memory tradeoff attacks)
- Proper statistical characteristics

It should be noted that the correlation attack is considered with its variants and derivatives such as fast correlation attack. The linear syndrome attack is a weak version of the fast correlation attack and need not be considered separately. The linear consistency attack is a divide & conquer technique, and it's enough to not use the structures that targeted the attack.

## III. OVERVIEW OF THE ALGORITHM

Our goal in this paper is not to define and express a new perspective on the design of stream cipher algorithms, but we want to design a new algorithm with high security based on system theory in such a way that the structures used in it are resistant to conventional attacks.

---
[1] Binary Symmetric Source

There are two major advantages to our proposed algorithm. First, we design a comprehensive algorithm (from A to Z) which describes every step, from the beginning of the message key generation until obtaining the final output keystream. The second advantage is the high security that has been proven in the security analysis section. In this section, threshold values have been obtained for all attacks that are considered as design criteria all of which are beyond the current computing power of the cryptanalysts. The structure of this algorithm is a little complex, which is also due to high security.

Generally, it is not so complicated to achieve high linear complexity and maximum period, as well as appropriate statistical features in the design of stream ciphers [4], but the resistance to known attacks and powerful new attacks (that have been invented in various studies) is not simple and can be the most important and difficult part of the design.

The base structure of the PALS algorithm is a clock-controlled combination generator with memory [5, 6]. We have achieved all the features according to design criteria with this structure. In Fig. 1, a simple combination generator is shown.

So far, using this combination and the proper design of the F function and LFSR's, we will achieve maximum period, high linear complexity, as well as the appropriate statistical properties [18]. However, several attacks such as algebraic and correlation attacks are still applicable against this structure. In the following, we will protect this structure against these attacks by using other tools.

*A. Correlation Immunity*

Combination generators are vulnerable to the various types of correlation attacks [18]. So, to make these attacks infeasible, the combining function F should have a high correlation-immunity order [17]. But, there exists a tradeoff between the correlation-immunity order and the algebraic degree of a Boolean function. We can use memory to conquest this tradeoff [7-9]. Using a bit of memory, correlation immunity is obtained, and we used this solution in the PALS algorithm.

In 2018, Deb et al. [6] proposed an LFER-based stream cipher which didn't consider this property and their output F function is formed only by three input bits without memory. Thus, their scheme is vulnerable against correlation attacks.

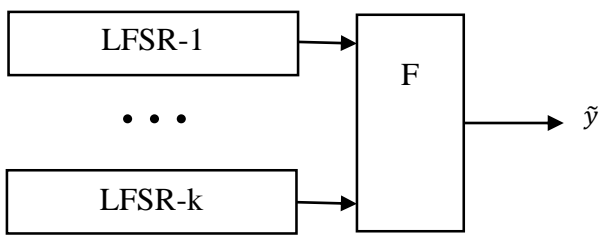

Fig. 1. Simple combination generator

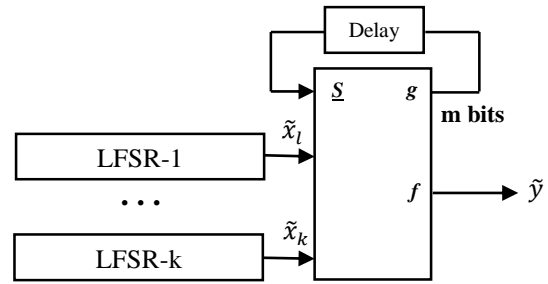

Fig. 2. Combination generator with memory

Fig. 2 shows a combination generator with memory, where k is the number of inputs to the function f and m is the number of inputs to memory.

In a linear attack, one can find a linear combination of at most m+1 consecutive output bits and a linear combination of maximum m+1 input vector sequences for an m-bits memory combination generator that correlate with each other. The main part of the linear attack [9], which is one of the most advanced correlation attacks, is the LSCA[2] algorithm for finding these linear transformations that have a complexity of order $2^{k+m}$. Considering the practical limitations, it is necessary to choose the value of m as small as possible and prevent the severe reduction of operational complexity of the linear attack with another trick. We will describe this trick in the next section.

Another important point to be mentioned here is how to design and select f and g functions. These functions should have the proper characteristics that are [10]:
1. The function f, the component functions of the vector function g, and also the linear combinations of these component functions should be balanced for any given fixed values of the input values $x_1, \ldots, x_k$.
2. The function f, the component functions of the vector function g, and also the linear combinations of these component functions should be balanced for any given fixed values of memory variables S.
3. The function f, the component functions of the vector function g, and also the linear combinations of these component functions should have high nonlinearity.

Finding the function f, which has the above features, is relatively simple. In other words, if we consider a stronger condition that the function f is an (n,1,t)-resilient function, in which t=max(m,k), then conditions 1 and 2 are easily satisfied for the function f. So, it is enough to find functions among the (n,1,t)-resilient function that has a high nonlinearity degree.

---

[2] Linear Sequential Circuit Approximation

## IV. BOOLEAN FUNCTIONS IN STREAM CIPHERS

Boolean functions are commonly used in stream ciphers to combine the output of LFSRs. These output sequences will be a pseudorandom sequence after a transient state [6].
*Definition 1*: The Boolean function of n variables f is a function of the set $F_2^n$ (all binary vectors with length n, such as $x=(x_1,…,x_n)$) to the field $F_2 = \{0,1\}$. It should be noted that finding the proper Boolean functions of n variables for cryptographic usage is very time-consuming for large n in exhaustive search (the number of Boolean functions of n variables is equal to $2^{2^n}$).
*Definition 2*: Assume that f is a Boolean function of n variables, then the following form is called the algebraic normal form (ANF) of the function f,

$$f(x_1,…,x_n) = a_0 \oplus (\bigoplus a_i x_i) \oplus (\bigoplus a_{ij} x_i x_j) \oplus … \oplus a_{12…n} x_1 … x_n \quad (1)$$

Where the coefficients $a_i$ are belong to the set $F_2=\{0,1\}$. Classical methods of designing Boolean functions can be divided into two categories:
- The first method does not pay attention to the algebraic degree and assumes that the number of variables and correlation immunity order is fixed.
- In the second method, the algebraic degree is taken into account; however, due to the Siegenthaler inequality [7], the maximum degree of algebraic value of an m-resilient function of n variables is equal to n-m-1.

In general, the necessary conditions to design Boolean functions in stream ciphers are balanced, high nonlinearity, high algebraic degree, and correlation immunity. The balanced and correlation-immune functions of order m are called m-resilient. With respect to the four mentioned properties [6,11-14], the best possible Boolean functions are designed in the PALS algorithm, i.e. they have a complex algebraic normal form with maximum nonlinearity, maximum algebraic degree, and at the same time, proper hardware implementation.

## V. DESCRIPTION OF THE PALS ALGORITHM

### A. Key Management

The main purpose of PALS is to construct an infinite (computationally) pseudorandom sequence, using a finite-length random sequence. Since the main key length of the encryption algorithm is 256 bits and much shorter than its initial state (1600 bits), this key must be extended by an appropriate method to obtain the initial vector. We also need to use a new key to encrypt each message, but we cannot change the main key for each message. Given these constraints, it is necessary to design a key generation algorithm that generates the initial vector by combining the main key of the system and a key called the message key.

It should be noted that in [6], there is no description about how to extend the main key bits and also, there is no message key in that. Thus, each new message is encrypted with the same key.

To do this, a session key is generated with the message key and the main key with the length of 256 bits. Then, the session key is expanded to 1600 bits (initial vector). The flowchart of initial vector generation is shown in Fig. 3.

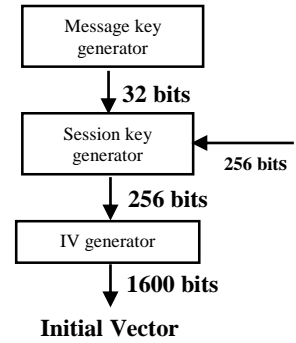

Fig. 3. Flowchart of Initial Vector generation

We also need to use a new key to encrypt each message, but we cannot change the main key for each message. Given these constraints, it is necessary to design a key generation algorithm that generates the initial vector by combining the main key of the system and a key called the message key.

It should be noted that in [6], there is no description about how to extend the main key bits. Also, there is no message key in that. Thus, each new message is encrypted with the same key.

To do this, a session key is generated with the message key and the main key with the length of 256 bits. Then, the session key is expanded to 1600 bits (initial vector). The flowchart of initial vector generation is shown in Fig. 3.

### B. Message Key Generator

The message key should be generated in such a way that it does not repeat during a period of main key change. If we assume that this algorithm is to be used 24 hours a day without interruption, and every second requires a 32 bits message key, after 4.25 years, a full period of the message key is generated and used and again returns to its initial state. Thus, the main key changeover period in this structure is up to 4.25 years. It's obvious that the main key should change much shorter than this period in cryptosystems.

But anyway, the calculations show that if the main keys are changed in four years intervals, we will not be worried about creating repetitive session keys. In the PALS algorithm, an LFSR with the length of 32 bits is used to generate a message key, whose feedback function is the following primitive polynomial:

$$C(x)=x^{32}+x^{29}+x^{24}+x^{23}+x^{21}+x^{19}+x^{17}+x^{16}+x^{14}+x^{13}+x^{11}+x^9+x^6+x^3+1 \quad (2)$$

### C. Session Key Generator

As previously mentioned, a scramble function must be designed to affect all the message key bits on the main key to generate the session key. With a fixed main key, the change of each message key bits should change each of the main key bits, with an average probability of 50% (Avalanche effect). This scramble function is created by a substitution-permutation network (SPN). To do this, the 32 bits message key is entered into a permutation box (P-Box) and a new 32 bits sequence is obtained. The resulting 32 bits are divided into 8 four-bit pieces, and each piece enters a substitution box (S-box$_{4\times4}$). The outputs of the S-boxes are concatenated respectively and produce a 32-bit sequence. In order to obtain good diffusion on all output bits by input changes, this operation must be repeated at least 5 times. As shown in Fig. 4, this iterative function is called Scram-5. Since the main key length of this algorithm is 256 bits, we need to use the Scram-5 function eight times to achieve a 256-bit sequence.

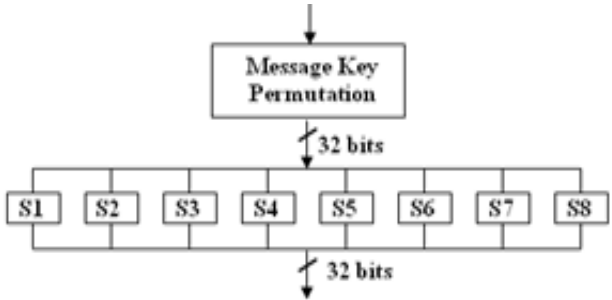

Fig. 4. Scramble function (Scram-5)

The session key is obtained by bitwise XORing the resulting 256-bit sequence with the main key (Fig. 5).

*D. Initial Vector Generator*

The length of the LFSRs used in the keystream generator is 1600 bits. So, we need a 1600-bit sequence for the initial state of them. As shown in Fig. 6, we used an LFSR of 256 bits, a primitive polynomial of degree 256 and four S-boxes to generate the initial vector.

The initial state of this LFSR is the 256-bit session key and generates eight bits at any clock. Four 8×8-bit S-boxes are embedded in the LFSR feedback function and one of them is selected and used to appropriate input-output diffusion.

The content of stage 128 and 129 are used to select one of the S-boxes (the sequence 00 will select the first S-box, 01 the second S-box, 10 the third S-box, and 11 the fourth S-box). After generating 320 bits (40 clocks), diffusion is achieved with 95% confidence. Therefore, it is necessary to discard the first 320-bit and the next 1600 bits are used as the initial state of the keystream generator.

In [6], the authors stated that all LFSRs are initially assigned with nonzero seed values. However, they didn't propose a procedure for that. In fact, since there is no message key in the proposed algorithm of these authors, it is not explained about its diffusion and the number of bits to be discarded initially.

*E. How to Initialize the Keystream Generator*

To initialize eight LFSRs used in the PALS keystream generator, the first 165 bits of the session key is divided into 163 three-bit sets as follows:

{1,2,3}, {2,3,4}, {3,4,5}, ..., {163,164,165}. Each set represents a decimal number from 0 to 7 and specifies one of the 8 LFSR's. For example, if the first set is 5, then the first bit of initial vector should be placed in the first stage of the sixth LFSR. Similarly, 163 bits of the initial vector is replaced in eight LFSRs. Then, the remaining 1437 bits of the initial vector is replaced in the empty stages of the LFSR's, respectively. This routine is only used for the first message key at the beginning of the communication. If the message key is to be sent repeatedly when sending a message, in order to synchronization between the receiver and the transmitter, for the next message keys the initial vector is XORed to the content of each LFSR from 1 to 8, respectively. Since in the algorithm of Deb et al. [6], the main key bits directly construct the initial state of the LFSRs, the necessary complexity is not achieved and the cryptanalysts can obtain the relationship between input and output bits. In other words, the initial locations of the bits are clear in the output sequence.

## VI. KEYSTREM GENERATOR

The main core of the keystream generator is contained 8 LFSRs with lengths (239, 163, 223, 181, 199, 173, 193, 229), which are relatively prime to each other. These LFSRs are clocked irregularly [17]. To do this, one of the 8×8-bit S-boxes is selected, which was used in the initial vector generator. To select the S-box, the output bits of the first, third, fifth and seventh LFSR are XORed and placed to the left. The output bits of the second, fourth, sixth and eighth LFSR are also XORed and placed to the right. As a result, a two-bit sequence is obtained, representing a binary number between 0 and 3, which chooses one of the four S-boxes. The input of the selected S-box is the output bits of the eight LFSRs. The 8-bit output of this S-box specifies which LFSRs should be clocked in each step (the least significant bit of the S-box output, corresponds to LFSR number eight, and the most significant bit corresponds to the first LFSR). Then, using the majority function, it is determined which of these LFSRs should be clocked. For example, if the S-box's output is 10111001, the first, third, fourth, fifth, and eighth LFSR are clocked and generate the new bit. The LFSRs that correspond to zero retain their previous value. It should be noted that since the number of LFSRs is even, all of the LFSRs will be clocked in the case of equality (four '0' and four '1'). Eight bits of different stages of these LFSRs are taken as input for eight nonlinear Boolean functions with nonlinearity of 6 and correlation-immunity order of 2. Each of these functions has 9 input variables and the value of the ninth variable of each one is one bit of 8-bit S-box's output. The least significant bit of the output goes to the function $F_1,\ldots,$ the most significant bit goes to the function $F_8$. The output of these eight functions enters to the 9-variable function g, which has the highest correlation immunity order.

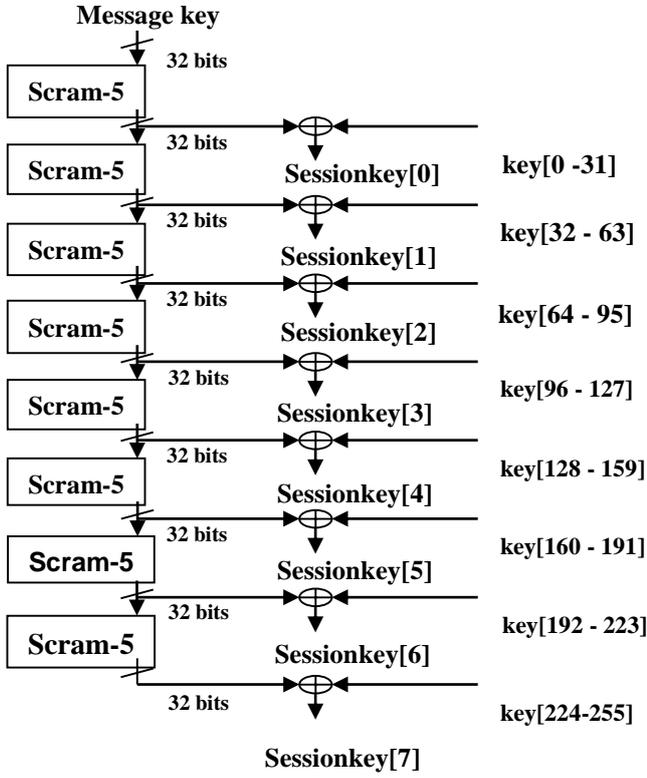

Fig. 5. Session key generator using Scram-5

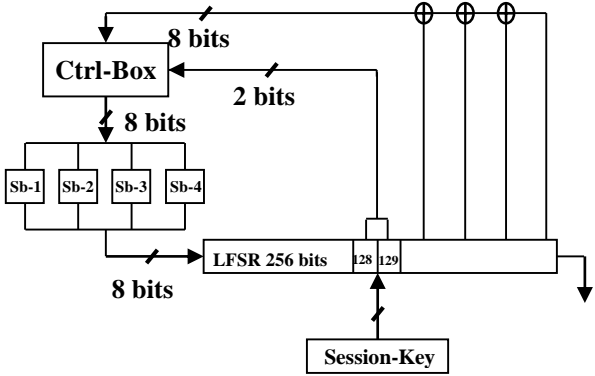

Fig. 6. Initial Vector Generator

The output of the previous step of the nonlinear function h is used as the ninth variable of the function g. The output of the function g is the keystream, which should be XORed with the message (plaintext). The LFSR's feedback polynomials are about half the length of their corresponding LFSR to resist against fast correlation attacks.

It should be noted that sparseness is an important issue in feedback functions that must be avoided, but in [6], the authors didn't mention it. The keystream generator is depicted in Fig. 7.

## VII. ANALYTICAL EVALUATION OF THE PALS

In this section, some result of analytical evaluation about PALS is described. In our proposed algorithm, each of the $2^n-1$ nonzero initial states of the non-singular LFSR of length n, produces an output sequence with maximum possible period $2^n-1$, because the feedback functions of LFSR's are primitive polynomials [16]. Meanwhile, the lengths of all LFSRs are chosen from the prime numbers.

So, considering the structure of the algorithm, its period is at least equal to the product of the individual period of them and is about $2^{1600}$. The large number of states also completely eliminates the threat of time-memory tradeoff attack on stream ciphers.

### A. Relation of Correlation Immunity with Nonlinearity of Boolean Functions

A non-linear function of n variables without memory has the correlation immunity of order m if the mutual information between the output variable Z and any subset of m input variable is equal to zero.

$$I(Z; x_{i_1}, \ldots, x_{i_m}) = 0 \ , \ 0 \leq i_1 \leq i_2 < \ldots < i_m \leq n-1 \quad (3)$$

There is always a tradeoff between the nonlinearity of the function f, i.e. k, with the correlation immunity order m. If n is the number of input variables of the function f, then $k+m \leq n$. When it is necessary that random output variable Z has uniform distribution (in cryptographic applications), this tradeoff will be:

$k+m \leq n$        *for* m=0 or m=n-1
$k+m \leq n-1$      *for* $1 \leq m \leq n-2$

Therefore, with the increase of correlation immunity order, the nonlinearity order of the function f and also the linear complexity of the generator decrease and vice versa. Using a trick, one can get a function that has the maximum order of correlation immunity and maximum nonlinearity order at the same time. In fact, this tradeoff can eliminate, using a single-bit memory in the input variables of the function. In addition, the nonlinearity of the function has no limitations and can be freely chosen. The output combiner of the PALS, i.e. the g and h functions are designed in this way. The algebraic form of these functions is as follows:

$h_i = X_1+X_2+X_5+X_5X_3+X_6X_4+X_7X_0+X_7X_1+X_7X_5+X_8X_0$
$+X_8X_2+X_8X_7X_0+X_8X_7X_1+X_8X_7X_3X_2+X_8X_7X_4X_2+X_8$
$X_7X_4X_3X_2+X_8X_7X_5X_2+X_8X_7X_5X_3X_2+X_8X_7X_5X_4X_2+$
$X_8X_7X_5X_4X_3X_2+X_8X_7X_6X_2+X_8X_7X_6X_3X_2+X_8X_7X_6$
$X_4+X_8X_7X_6X_4X_2+X_8X_7X_6X_4X_3+X_8X_7X_6X_4X_3X_2+X_8$
$X_7X_6X_5+X_8X_7X_6X_5X_2+X_8X_7X_6X_5X_3+X_8X_7X_6X_5X_3$
$X_2+X_8X_7X_6X_5X_4+X_8X_7X_6X_5X_4X_2+X_8X_7X_6X_5X_4X_3+$
$X_8X_7X_6X_5X_4X_3X_2$                                                 (4)

$$g = X_0+X_1+X_2+X_3+X_4+X_5+X_6+X_7+h_{i-1} \quad (5)$$

To increase the speed of software implementation, the truth table of the above functions can be used in the form of lookup tables.

### B. Correlation Immunity in PALS

To show the strength of the algorithm against analytic correlation attacks, we first need to calculate the total number of possible keys to construct the initial state of the algorithm. To do this, we need to compute the number of available feedback polynomials and the number of possible initial states for each LFSR. The number of feedback polynomials of degree $L_i$ is:

$$U_i = [\varphi(2^{L_i}-1)]/L_i \quad (6)$$

So the total number of the keys is equal to:

$$\Pi U_i \times (2^{L_i}-1) \qquad (7)$$

Assuming that the initial state and polynomial feedback of each LFSR can be calculated separately, the total number of keys will be equal to:

$$\Sigma U_i \times (2^{L_i}-1) \qquad (8)$$

As a result, the total number of the PALS's keys is equal to:

$$2^{238} \times (2^{239}-1) + \ldots + 2^{228} \times (2^{229}-1) > 2^{477} \qquad (9)$$

Therefore, a simple correlation attack against the algorithm is impossible (computationally) and PALS has the practical security against this attack. In general, any kind of fast correlation attack against the PALS is impractical due to the clock control structure of the algorithm, the use of dense feedback polynomials, the use of correlation immune and nonlinear functions and the large 1600 bits initial state of the algorithm.

### C. Algebraic Attack

An algebraic attack attempts to construct the keystream of the algorithm by establishing a relationship between each output and input bit of the algorithm. Therefore, any simple relation between the input and output bit of the algorithm should be eliminated [15]. In the PALS algorithm, the method of initialization (first 163 bits of LFSRs), irregular clock structure and the use of the output of nonlinear functions instead of direct use of the LFSRs outputs on the input of the function g, creates necessary complexity, and the attackers cannot actually find a relationship between input and output bits, because the initial location of the bits, LFSR's number, stage and clock of LFSRs are unclear in output sequence.

### D. Times-Memory Tradeoff Attack

*This attack is based on the birthday paradox.*

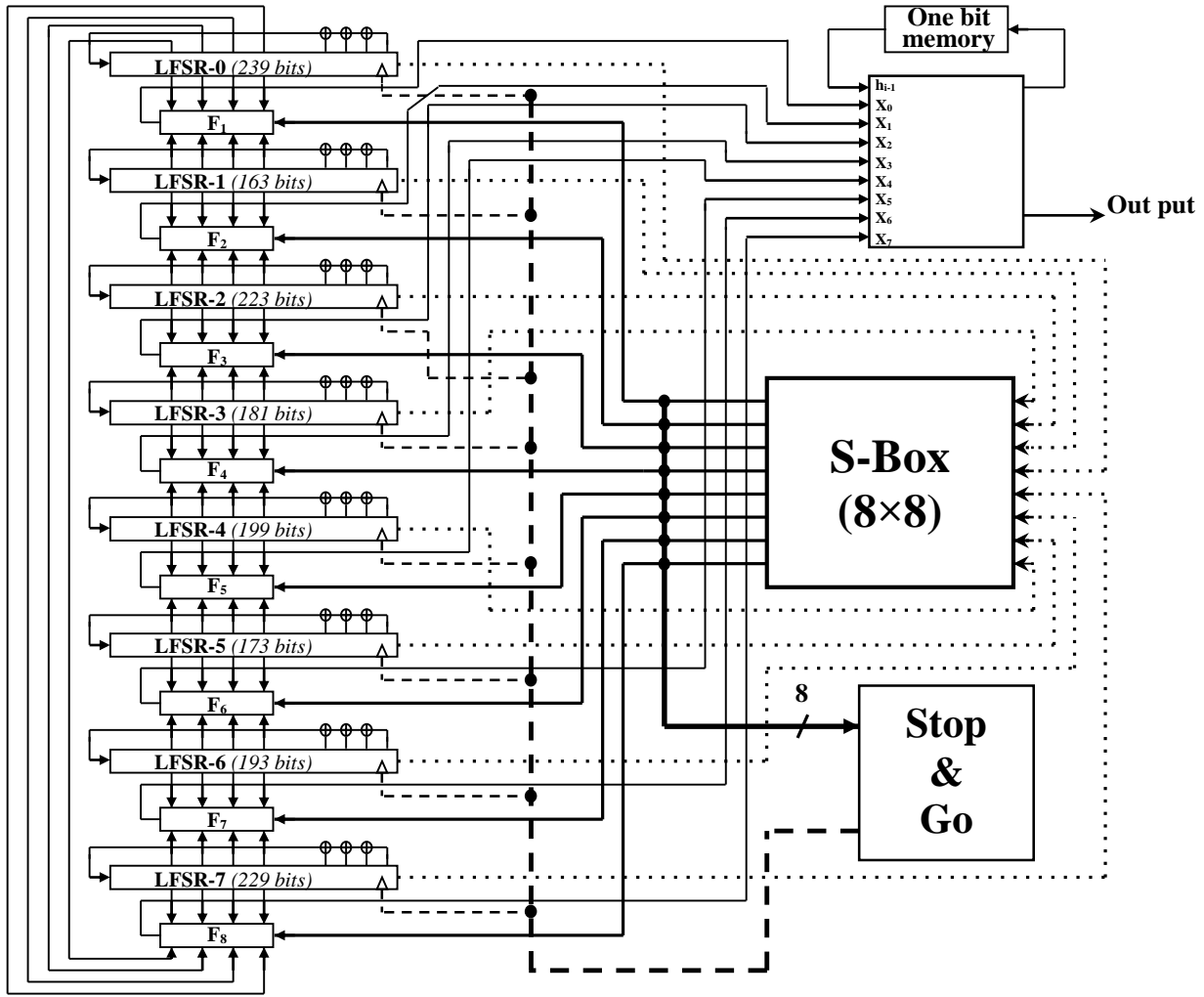

Fig. 7. Keystream Generator

The birthday paradox is often used to show how much data should be stored in memory to achieve a certain probability of success of the attack. By choosing n random bits as the initial state, it produces n bits of keystream and compares it with the content which was previously stored in memory ($2^m$ n-bit output sequence, m=n/2). If a match occurs, a state of the system is obtained. In the event of the failure at this stage, a new n random bit is used and this procedure is repeated. The memory and computational order of this attack are obtained from the following relationships:

$$T=O(n+m^2)(2^m+2^{n-m}) \qquad (10)$$

$$M = O(n+m)(2^m) \quad (11)$$

The initial state of the PALS algorithm is 1600 bits and the value of m is equal to 800. So, the memory and computational order to implement the time-Memory tradeoff attack against the algorithm is equal to:

$$T = O(1600+800^2)(2^{800}+2^{1600-800}) = 2^{820} \quad (12)$$

$$M = O(1600+800)2^{800}) = 2^{810} \quad (13)$$

These amounts of memory and computational order are too large enough and the attack is impractical.

*E. Divide & Conquer Attack*

This attack is based on the idea of time-memory tradeoff attack, and according to the above illustrations, its preprocessing time, processing volume, and memory requirements are all larger than $2^{500}$ and obviously the divide & conquer attack is also impractical against the proposed algorithm.

*F. Distinguishing Attack*

This attack attempts to obtain the initial state of the algorithm by taking a long sequence of output bits. Since the linear complexity of this algorithm is greater than $10^{480}$, assuming that the algorithm produces $10^{12}$ bits per second, after a period of 1000 years, the attackers will only have $10^{22}$ output bits. Therefore, it is shown that this attack is not applicable to PALS.

*G. AIDA/cube Attacks*

AIDA/cube attacks [19] are generic key-recovery attacks that can be used to encryption algorithms without the need to know the internal structure of the algorithm. An important requirement is that the output from the generator can be represented as a low-degree decomposition multivariate polynomial in the algebraic normal form in the key and the plaintext. We assume the attacker is allowed to query the master polynomial (that is, a chosen-plaintext, chosen-IV setting) of its choice and achieve the resulting bit from the master polynomial. This way, the attacker achieves a system of polynomial equations in terms of secret variables only. The ultimate goal of the attack is to solve this system of equations, which reveals the key variables [20].

But the AIDA/cube attack is applicable on the algorithm, if the output keystream can be represented by a low degree multivariate polynomial. In other words, this attack is successful when applied to random polynomials of degree d over n secret variables, whenever the number m of public variables exceeds $d + \log_n d$. The complexity is $2^{d-1}n + n^2$ bit operations, which is polynomial in n and low when d is small. The polynomials in PALS have degrees from 163 to 239. For instance, the complexity of this attack is at least $2^{162}$ bit operations that virtually make it impossible to implement in a reasonable time.

## VIII. CONCLUSIONS

Due to the wide use of stream ciphers in various applications, in this paper, we proposed a new stream cipher named PALS, which can be taken as a clock-controlled one, but with a new mechanism of altering steps. The Length of the main key is equal to 256 bits, but it can be 512 or 1024 bits with a little change in initial vector generator.

The most important criteria considered in designing the proposed algorithm are the maximum period, high linear complexity, resistance to known attacks and good statistical characteristics. The base structure of the PALS algorithm is a clock-controlled combination generator with memory and we obtained all the features according to design criteria with this structure.

Our proposed algorithm is designed based on system theory and very similar to a BSS generator, so its keystream is like the perfectly random sequences and resistant to conventional attacks such as algebraic attack, time-memory tradeoff attack, divide & conquer attack, distinguishing attack, AIDA/cube attack, and various types of correlation attacks.

We achieved a good avalanche effect (message key on the main key), using the substitution-permutation boxes repeatedly in the session key generator. Also, the use of Boolean functions (with appropriate cryptographic properties) increased algebraic degree and nonlinearity of the algorithm adequately.

The main advantage of the proposed algorithm is the high security that has been proven in the security analysis section. In this section, threshold values have been obtained for all attacks that are considered as design criteria all of which are beyond the current computing power of the cryptanalysts.

Due to the proper security features of the PALS, it can be used in many applications, especially in financial cryptography. As we evaluated PALS is secure against of well-known attacks. However, the extensive security analysis of any new cipher requires a lot of efforts from many researchers. We thus invite and encourage the readers to analyze the security of PALS.